# Lumped-element SNAIL parametric amplifier with two-pole matching network

D. Moskaleva,[1,2] N. Smirnov,[1] D. Moskalev,[1] A. Ivanov,[1] A. Matanin,[1]
D. Baklykov,[1] M. Teleganov,[1] V. Polozov,[1] V. Echeistov,[1] E. Malevannaya,[1]
I. Korobenko.[1] A. Kuguk,[1] G. Nikerov,[1] J. Agafonova,[1] and I. Rodionov[1,2,a)]

[1]*FMN Laboratory, Bauman Moscow State Technical University, Moscow, 105005, Russia*
[2]*Dukhov Automatics Research Institute, VNIIA, Moscow, 127030, Russia*

**a) Author to whom correspondence should be addressed**: irodionov@bmstu.ru

Broadband impedance-matched Josephson parametric amplifiers are key components for high-fidelity single-shot multi-qubit readout. Nowadays, several types of impedance matched parametric amplifiers have been proposed: the first is an impedance-matched parametric amplifier based on a Klopfenstein taper, and the second is the impedance-matched parametric amplifier based on auxiliary resonators. Here, we present the quantum-limited 3-wave-mixing lumped-element SNAIL parametric amplifier with two-pole impedance matching transformer. A two-pole Chebyshev matching network with shunted resonators based on parallel-plate capacitors and superconducting planar coil. Operating in a flux-pumped mode, we experimentally demonstrate an average gain of 15 dB across a 600 MHz bandwidth, along with an average saturation power of –107 dBm and quantum-limited noise temperature.

Josephson parametric amplifiers have become key components in quantum information processing[1] as multi-qubit readout fidelity can limit overall quantum algorithms accuracy[2]. The dynamic range of parametric amplifiers and their noise temperature mostly define the readout fidelity. State-of-the-art quantum-limited parametric amplifiers realize nondestructive high-fidelity single-shot readout of superconducting qubit[3-6] and multiplexed qubits readout[7-9]. In addition, broadband Josephson parametric amplifiers are key elements in single-photon power measurements of microwave[10-12] and optical[13] signals, ultrahigh efficient multi-resonator quantum memory readout[14], fast high-quality resonators characterization[15], quantum metrology[16], and dark matter research[17]. At present, traveling wave parametric amplifiers (TWPA) demonstrate the highest dynamic range, but their noise temperature is much higher than the quantum limit, which limits the quantum efficiency[18-19].

Impedance-matched Josephson parametric amplifiers (IMPA) are one of the most perspective quantum-limited cryogenic amplifiers. The idea of using impedance matching for parametric amplifiers was first demonstrated by the J. Martinis group in 2014[20]. Using a Klopfenstein taper they changed the impedance slowly from 50 Ω (room temperature electronic impedance) to load impedance $Z_L$ = 15 Ω. Later, this approach was realized by several scientific groups[21-23]. IMPA based on the Klopfenstein taper demonstrated a 15 dB gain across bandwidth from 300 to 700 MHz, along with an average saturation power of -110 dBm. In 2015, the IMPA based on auxiliary resonators[24] was first demonstrated. The auxiliary resonators were realized on a PCB plate as a two-pole Butterworth prototype. Later, impedance-matched parametric amplifiers based on the auxiliary resonators were realized on-chip[25-28]. It is characterized by the gain of at least 15 dB, 200…600 MHz bandwidth and the saturation power about -100…-90 dBm. The best result for IMPA dynamic range was demonstrated in Ref. 29, where a three-pole Chebyshev matching network was realized by two lumped element resonators and one a transmission line resonator.

One of the key disadvantages of the proposed solutions is distributed auxiliary resonators realization. This significantly limits the range of characteristic impedances (from 15 Ω to 150 Ω), and increases devices footprint. Usually, IMPA nonlinear resonator have the characteristic impedance lower than 15 Ω (by reducing the impedance of the auxiliary resonator one can limit the reflections). Moreover, in the case of Klopfenstein taper-based amplifiers, it is necessary to fabricate a complex

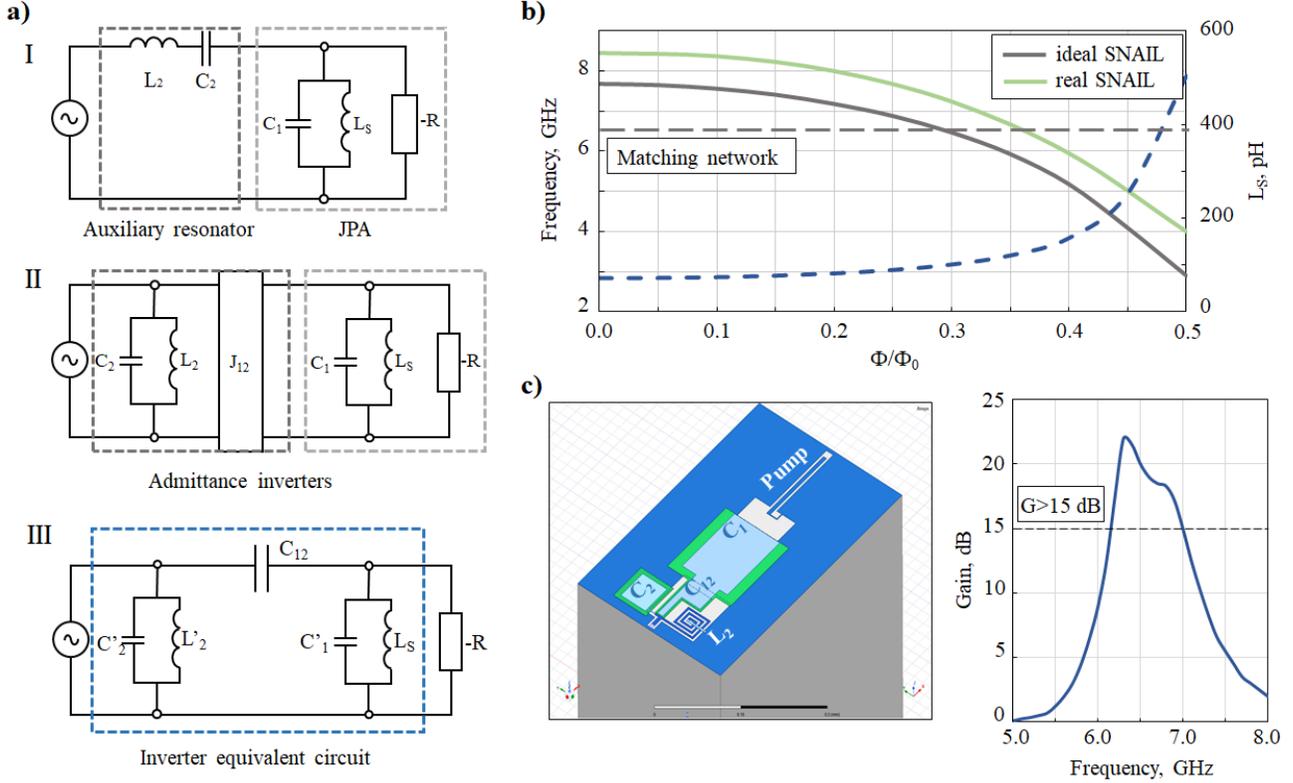

FIG. 1 Simulation results of IMPA with the two-pole Chebyshev matching network. (a) IMPA electric circuit transformation by the admittance inverters integrating. (b) Simulation of the IMPA tunable range for the asymmetry coefficient $a = 0.25$. Solid grey line indicates a simulation of the tunable range for ideal SNAIL ($L_{j1}=L_{j2}=L_{j3}$). And green line indicates the simulation for real SNAIL including fabrication fluctuations ($L_{j1}=L_{j3}\neq L_{j2}$) (c) Simulation of an ideal IMPA resulting gain using Ansys HFSS

transmission line consisting of hundreds of variable capacitors. This problem can be solved using lumped-element resonators[29-30].

In this work, we present a broadband lumped-element impedance-matched parametric amplifier consisting of a single superconducting nonlinear asymmetric inductive element (SNAIL) and the on-chip two-pole Chebyshev matching network. Compared with Ref. 29, the proposed device doesn't have distributed elements, and the two-pole Chebyshev filter is easier to design than the three-pole filter without scarifying the gain and bandwidth. Our amplifier operates in 3-wave-mixing reflection mode when the input signal reflects off, generating an amplified output signal with the gain of more than 15 dB and idler tone. We used a flux line to tune the SNAIL operating point. To improve the dynamic range of our parametric amplifier, we transform the environmental impedance, increasing coupling, lowering $Q = Z\omega C = Z/\omega L$ and using the single SNAIL proposed in previous works[31-33]. We engineered the IMPA with the average gain of 15 dB across the 600 MHz bandwidth, along with the average saturation power of –107 dBm, and quantum-limited noise performance.

The impedance-matched parametric amplifier consists of a JPA and the on-chip two-pole impedance transformer. In our case, JPA is the LC oscillator realized by the parallel connection of the single SNAIL with an inductance $L_S$ and parallel plate capacitor with capacitance $C_1 = 5.25$ pF. The IMPA central resonance frequency is defined as $\omega_0 = 1/\sqrt{L_S C_1}$, where $L_S$ define as follows:

$$L_s(\varphi_{ext}) = L_J/c_2(\varphi_{ext}), \qquad (1)$$

where $L_j$ is the Josephson inductance of large junction in the loop, $c_2 = \alpha \cos \varphi_{min} + \frac{1}{3}\cos\left(\frac{\varphi_{min}-\varphi_{ext}}{3}\right)$ is the flux-tunable constant, $\alpha = \frac{L_{j(large)}}{L_{j(small)}}$ is the cell asymmetry, $\varphi_{ext} = 2\pi\Phi/\Phi_0$ is the magnetic flux quantum and $\varphi_{min}$ defined as $c_1 \equiv \alpha \sin \varphi_{min} + \sin \frac{\varphi_{min}-\varphi_{ext}}{3} = 0$. We choose SNAIL as a nonlinear element because its 3rd and 4th order (Kerr) nonlinearities depends on an external magnetic flux and possible to find the optimal operating point with negative Kerr nonlinearity while third-order nonlinearity depicted a 3-wave mixing process will be maximized. Limiting Kerr nonlinearity improves the IMPA saturation power and maximizes the gain[33].

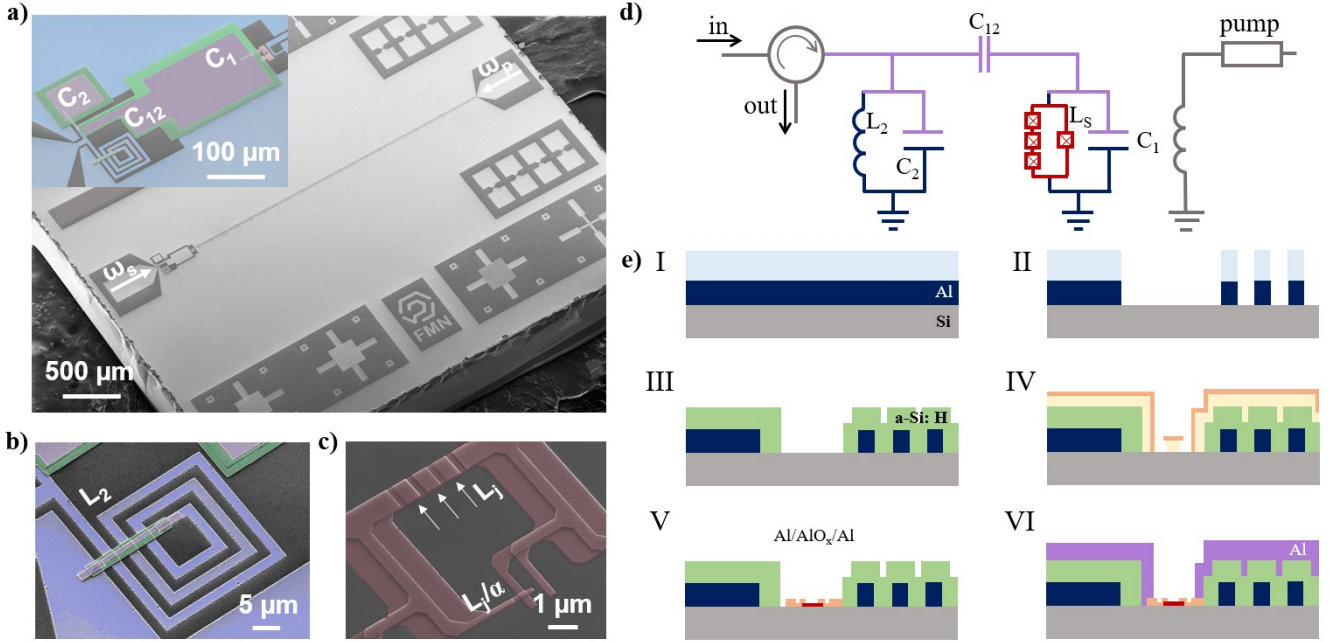

FIG. 2 IMPA with two-pole Chebyshev matching network. (a) Micrograph of the fabricated device with two limped-element resonators. (b) Higher-magnification micrograph of planar coil. (c) Higher-magnification micrograph of SNAIL, which consists of an array of 3 large Josephson junctions in a loop with one smaller junction (red). (d) Circuit representation of the device. (e) IMPA fabrication steps.

For the impedance transformer simulation, the negative-resistance prototype method proposed in Ref. 34 was utilized. We used the 2nd-order Chebyshev prototype (equal-ripple) to provide a multi-peak gain for the impedance transformer circuit. We chose prototype[34] with minimum gain $G_{min}$ = 20 dB, passband ripple 0.5 dB, and prototype coefficients are $g_1$ = 0.5, $g_2$ = 0.24 and $g_3$ = 1.22. The gain is defined as the ratio of the reflected power dissipated in the load resistance to the power available from the generator and can be described as follows:

$$G = |S_{11}|^2 = |\Gamma(\omega)|^2 = \frac{P_L(\omega) - 1}{P_L(\omega)}, \quad (2)$$

where $P_L(\omega) = 1 + k^2 T_N^2(\omega)$ is the Chebyshev power loss function and $T_N = 2\omega^2 - 1$ is the Chebyshev polynomial. On the other hand, the reflection coefficient $\Gamma(\omega)$ depends on the input impedance of IMPA circuit ($Z_{in}$): $\Gamma(\omega) = (Z_{in} - Z_0)/(Z_{in} + Z_0)$, where $Z_{in}$ is the input impedance of IMPA. The bandwidth improvement of this IMPA is defined by fractional bandwidth $w$, which determine as $w = \frac{\omega_2 - \omega_1}{\omega_0}$, where $\omega_0 = \sqrt{\omega_1 \omega_2}$ is the central resonance frequency and R is the negative resistance of parametric amplifier.

FIG. 1(a) I shows the electrical circuit of the parametric amplifier formed by the negative-resistance prototype method. It consists of an alternating array of parallel and series resonators. However, resonators with characteristic impedance about $Z_{series}$ > 150 Ω for series connection resonators and $Z_{shunt}$ < 15 Ω for parallel connection resonators cannot be realized by microstrip or coplanar transmission lines. That's why we used an admittance invertor for electrical circuit transformation[35] [see FIG. 1(a) I…III]. In this study, the admittance inverter is a capacitance element that integrated into the electrical circuit, and it changes the resonator connection scheme. The inverter coupling capacitance depends on the inverter constant and the central resonance frequency of parametric amplifier $\omega_0$ Eq. (3).

$$C_{ij} = \frac{J_{ij}}{\omega_0}, \quad (3)$$

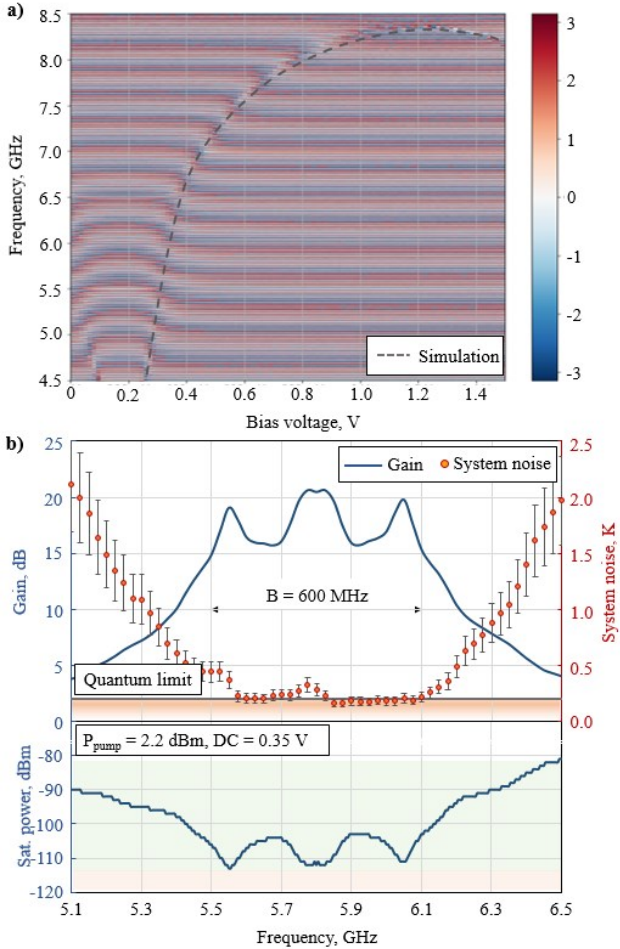

FIG. 3. Measured device performance operated at 10 mK. (a) Experimentally measured unwrapped phase response of the IMPA vs magnetic flux bias. Dashed grey line indicates a simulation of the tunable range of the IMPA including fabrication imperfections. (b) We demonstrate the amplifier performance at center frequency $\omega_a/2\pi$ = 5.8 GHz. The IMPA provides an input saturation power of -107 dBm. The device shows a gain above 15 dB and operational bandwidth of nearly 600 MHz. We define the quantum limit as one photon $\hbar\omega$ of total system noise at the input of the amplifier.

where $J_{ij} = \frac{w}{\sqrt{g_i g_j Z_i Z_j}}$ is the inverter constant, $g_{ij}$ are the Chebyshev prototype constants and $Z_{ij}$ are the impedance of the resonators.

We used the admittance inverter and resonators parallel connection because series resonator has a significant parasitic capacitance negatively influenced on the gain performance. Detailed calculation of the main IMPA parameters presented in supplementary material.

We additionally simulate the IMPA tunable range [see FIG. 1(b)], which depends on the asymmetry coefficient (α) of the single SNAIL. We demonstrate that the differences between series-connected josephson junctions in the SNAIL loop that arise due to fabrication imperfections affect the IMPA tunable range. Also, we used Ansys HFSS to obtain the optimal gain profile of the final IMPA design [see FIG. 1(c)].

FIG. 2 shows scanning electron microscopy images of the impedance-matched Josephson parametric amplifier and its fabrication process. Resonator 1 is formed by a capacitor $C_1$ = 5.25 pF that shunts the single SNAIL with inductance $L_S$. The SNAIL consists of an array of 3 large Josephson junctions (with Josephson inductance $L_J$ = 40 pH) in a loop and one smaller junction (with inductance $L_J/\alpha$) which has the asymmetry α = 0.25 [see FIG. 2(c)]. We choose the asymmetry coefficient based on experimental dependence on the average parametric amplifier gain from α (see supplementary materials). The device was designed for the center frequency of $\omega_0/2\pi$ = 6.5 GHz and fractional bandwidth of $\Delta\omega/2\pi$ = 500 MHz. Resonator 2 was implemented as the lumped-element parallel LC resonator with $Z_2$ = 12.7 Ω [see FIG. 2(d)]. The coupling was realized by capacitors $C_{12}$ = 0.75 pF. To prevent microwave-frequency slot modes, Al cross-overs were fabricated with the same dielectric as the capacitors placed on the pump port every 250 μm.

The IMPA fabrication process includes four main steps: (I, II) ground layer patterning (planar spiral coils, bottom capacitor electrodes, etc.), (III) dielectric layer deposition, (IV, V) SNAIL e-beam lithography, double-angle evaporation and lift-off, and (VI) evaporation of the top electrodes of the capacitors [see FIG. 2 (e)]. Before ground layer evaporation, silicon substrates were cleaned in Piranha solution at 80°C, and then 100-nm aluminum was deposited by electron-beam evaporation method. We used dry etching in Ar/Cl₂ for planar coils and bottom capacitor electrodes forming. A 350 nm thick hydrogen-saturated amorphous silicon (a-Si:H) layer was used as the dielectric layer in the capacitors. The PECVD method was used for dielectric stack deposition, and dry etch in CF₃/Ar was implemented for a-Si: H etch. Then, a bilayer mask was spin-coated onto the substrate, which consisted of 500 nm MMA (methyl methacrylate) and 150 nm AR-P (CSAR). The SNAIL is patterned using direct 50 kV e-beam lithography. Al/AlOₓ/Al Josephson junctions are e-beam shadow-evaporated in a single vacuum cycle[36-37]. Lift-off was performed in a bath of N-methyl-2-pyrrolidone with sonication at 80°C for 3 hours and was rinsed in IPA with ultrasonication. The capacitor top electrodes were formed by a 600 nm thick Al.

Finally, we experimentally tested our amplifier in a dilution refrigerator with a base temperature below 10 mK. The cryogenic characterization has performed in a reflection mode with a circulator connected in series. The flux bias of the SNAIL loops was controlled by the flux line. Here, we define the operational bandwidth as the frequency range in which the gain is greater than 15 dB, and the noise temperature corresponds to the standard quantum limit. First, we determined the IMPA characteristic frequency by applying flux to SNAIL loop through a flux line. The resonant frequency in our case can be tuned from 3.9 to 8.4 GHz [see FIG. 3(a)]. Second, we chose the flux-bias operating point $\Phi_{DC} = 0.42\Phi_0$ and measure the gain response. We focused on areas with negative Kerr for SNAIL IMPA characterization.

FIG. 3(b) shows the best gain profile obtained at central frequency $\omega_{JPA}$ = 5.8 GHz with a bandwidth of 600 MHz. This profile was obtained at the source pump power of $P_{pump}$ = 2.2 dBm and flux line voltage of 0.35V. The measuring device operates in 3-wave mixing mode. The saturation power of the amplifier is defined as a value of the input signal power at which the gain is decreased by 1 dB. The saturation power was measured at the same center frequency $\omega_{JPA}$ = 5.8 GHz. We observed the saturation power of −105…−110 dBm in the bandwidth of 600 MHz with the gain above 15 dB [see FIG. 3(b)]. We observed some a lack of correspondence between simulation and measurement results. We explained this phenomenon in the supplementary material.

For the noise temperature measurements, we used the method described in Ref. [20]. We calibrated the noise of the HEMT following our IMPA, and then calculated the system noise using SNR improvement method. The estimated IMPA noise temperature is consistent with the near-quantum-limited operation [see FIG. 3(b)].

In summary, we have designed, fabricated, and characterized the impedance-matched Josephson parametric amplifier based on the single SNAIL and the lumped-element impedance transformer based on the two-pole Chebyshev matching network. The proposed device has a stable gain profile, wide bandwidth, and the standard microelectronic fabrication process. We demonstrated the average gain of 15 dB across a 600 MHz bandwidth at the central resonance frequency of 5.8 GHz. The noise temperature was estimated to be close to the standard quantum limit with saturation power from −105 dBm to −110 dBm over the bandwidth. There is no need in λ/4 impedance transformer in the proposed design. The characteristic impedance of the auxiliary resonator (12.7 Ω) is quite difficult to realize by distributed transmission lines. Moreover, the lumped-element IMPA realization is feasible for further improvements by increasing of auxiliary resonators and easier integration of on-chip filters and circulators. Our amplifier is suitable for quantum information processing, such as multi-qubit readout, quantum vacuum measurements, and microwave photon detection.

## AUTHOR DECLARATIONS

**Conflict of Interest**
The authors have no conflicts to disclose.

## AUTHOR CONTRIBUTIONS

D. Moskaleva, N. Smirnov and D. Moskalev contributed equally to this work.

**Daria Moskaleva:** Conceptualization (equal); Formal analysis (lead); Methodology (equal); Investigation (lead); Writing – original draft (lead). **Nikita Smirnov:** Conceptualization (equal); Formal analysis (lead); Investigation (lead); Writing – review and editing (equal). **Dmitry Moskalev:** Conceptualization (equal); Formal analysis (lead); Methodology (equal); Investigation (lead); Writing – review and editing (equal). **Anton Ivanov:** Formal analysis (equal); Investigation (equal); Writing – review and editing (equal). **Alexey Matanin:** Formal analysis (equal); Investigation (supporting); Writing – review and editing (supporting). **Dmitry Baklykov:** Investigation (supporting); **Maksim Teleganov:** Investigation (supporting); **Victor Polozov:** Investigation (supporting); **Vladimir Echeistov:** Investigation (supporting); **Elizaveta Malevannaya:** Investigation (supporting); **Igor Korobenko:** Investigation (supporting); **Alexey Kuguk:** Investigation (supporting); **Georgiy Nikerov:** Investigation (supporting); **Julia Agafonova:** Investigation (supporting); **Ilya Rodionov:** Project administration (lead); Conceptualization (lead); Formal analysis (equal); Writing – review and editing (equal).

## DATA AVAILABILITY
The data that support the findings of this study are available within the article and its supplementary material.

# Supplementary material for

# Lumped-element SNAIL parametric amplifier with two-pole matching network

D. Moskaleva,[1,2,b)] N. Smirnov,[1,b)] D. Moskalev,[1,b)] A. Matanin,[1] A. Ivanov,[1] D. Baklykov,[1] M. Teleganov,[1] V. Polozov,[1] V. Echeistov,[1] E. Malevannaya,[1] I. Korobenko,[1] A. Kuguk,[1] G. Nikerov,[1] J. Agafonova[1] and I. Rodionov[1,2, a)]

[1] FMN Laboratory, Bauman Moscow State Technical University, Moscow, 105005, Russia

[2] Dukhov Automatics Research Institute, VNIIA, Moscow, 127030, Russia

---

a) Author to whom correspondence should be addressed: irodionov@bmstu.ru

b) D. Moskaleva, N. Smirnov and D. Moskalev contributed equally to this work.

## CALCULATION OF IMPA CIRCUIT MAIN PARAMETERS

We designed the IMPA to have a center resonance frequency of $\omega_0/2\pi = 6.5$ GHz. The circuit of the parametric amplifier is shown in FIG. 2(d) in the main text. Resonator 1 in FIG. 2(d), having a characteristic impedance $Z_1$, consists of a single superconducting nonlinear asymmetric inductance element (SNAIL) with the critical current of a large josephson junction in a loop $I_c = 7.9$ uA and parallel-plate capacitor $C_1$.

Resonator 2 with characteristic impedances $Z_2$ is the lumped-element parallel LC resonator, and its resonance frequency is equal to the parametric amplifier resonance frequency. The resonators are interconnected via the admittance inverter $J_{12}$. We start from the Chebyshev prototype coefficients choosing and calculating the fractional bandwidth $w = 0.12$ [1]. We fix the impedance of the 1st resonator $Z_1 = \omega_0 L_s = 4.22$ Ω. Then, we calculate the impedance of 2nd resonator $Z_2 = 12.7$ Ω and the value of the admittance inverter [2] using Eq. (S1).

$$J_{12} = \frac{w}{\sqrt{g_1 g_2 Z_1 Z_2}}, \tag{S1}$$

where the Chebyshev prototype coefficients $g_{ij}$ described in the main text. The value of the admittance inverter is $J_{12} = 0.045 \, \Omega^{-1}$. We implement the admittance inverter disposed between two LC resonators, using coupling capacitance $C_{12} = \frac{J_{12}}{\omega_0} = 0.75 \, pF$.

Next, we calculate all IMPA circuit parameters. Capacitors $C_1$ and $C_2$ are calculate using Eq. (S2).

$$C_i = \frac{1}{Z_i \omega_0} - C_{ij} \tag{S2}$$

The resonator components are $L_2 = 0.32$ nH, $C_2 = 1.13$ pF, and $C_1 = 5.25$ pF. Finally, we simulate the gain profile for our IMPA. The resulting gain of parametric amplifier define by Eq. (S3).

$$G = |S_{11}|^2 = |\Gamma(\omega)|^2 = \frac{Z_2 - Z_0}{Z_2 + Z_0}, \tag{S3}$$

where, $Z_2 = 1/(1/Z_1 + j\omega_0 C_2 - j/\omega_0 L_2)$, $Z_1 = 1/(1/R + j\omega_0 C_1 - j/\omega_0 L_1) - j/\omega_0 C_{12}$.

## SUPERCONDUCTING NONLINEAR ASYMMETRIC INDUCTIVE ELEMENT (SNAIL)

The SNAIL consists of an array of 3 large Josephson junctions (with the Josephson inductance $L_J = 40$ pH) in a loop with one smaller junction (with inductance $L_J/\alpha$) which has an asymmetry $\alpha = 0.25$. The potential energy of SNAIL [3] can be described by Eq. (S4).

$$U_s(\varphi) = E_J\left(\frac{c_2}{2!}(\varphi - \varphi_{min})^2 + \frac{c_3}{3!}(\varphi - \varphi_{min})^3 + \frac{c_4}{4!}(\varphi - \varphi_{min})^4 + \cdots\right), \tag{S4}$$

where $c_2 = \alpha \cos\varphi_{min} + \frac{1}{3}\cos\left(\frac{\varphi_{min} - \varphi_{ext}}{3}\right)$ described the parametric amplifier tunable range, $c_3 = -\alpha \sin\varphi_{min} + \frac{1}{9}\sin\frac{\varphi_{min} - \varphi_{ext}}{3}$ described the maximum gain of the amplifier, $c_4 = -\alpha \cos\varphi_{min} - \frac{1}{27}\cos\left(\frac{\varphi_{min} - \varphi_{ext}}{3}\right)$ described the saturation power, $\alpha$ is the asymmetry coefficient, $E_j = \frac{\Phi_0 I_c}{2\pi}$ is the Josephson energy with inductance $L_J$.

The maximal asymmetry coefficient defined as $\alpha \leq 1/n$, and for the proposed device equal $\alpha = 0.33$. The increase in the asymmetry coefficient increases the third-order nonlinearity, which improved the maximal gain. We demonstrate experimental study of the influence of the asymmetry coefficient on the resulting gain profile. We used our previously developed IMPA circuit based on the SNAIL array [4]. We fabricated and characterized the three IMPA devices with different asymmetry coefficient only ($\alpha$ = 0.10, 0.20 and 0.25). FIG. S1 shows the experimental gain profile for these devices. We demonstrate increasing of the maximal gain without scarifying the bandwidth with increasing the asymmetry coefficient.

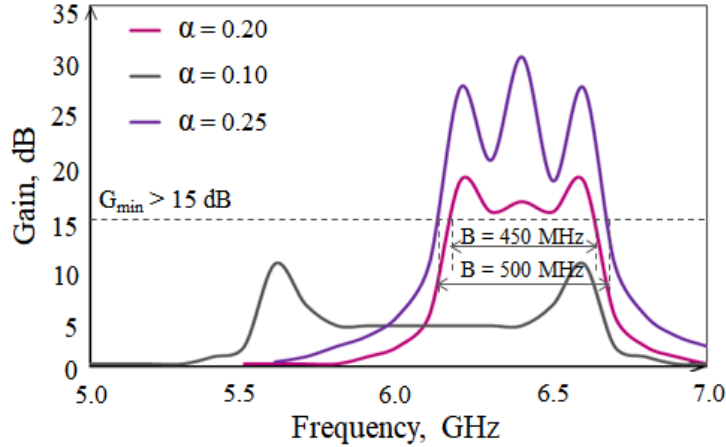

FIG. S1. Experimental dependence of the gain from coefficient asymmetry values α

**EXPERIMENTAL SETUP**

The experimental setup is shown in FIG. S2. The IMPA was packaged in a copper sample holder with aluminum and cryoperm shields mounted on a mixing plate of a dilution refrigerator at a temperature below 10 mK. The transmission measurements via the circulator were performed using a Vector Network Analyzer (VNA). The input line was attenuated at various stages (total 70 dB). The output line included the HEMT amplifier at 4 K and the room-temperature low-noise amplifier (LNA). The pump line was attenuated at 4 K and 100 mK stages with total 30 dB attenuation and included high-pass filter (HPF). The global magnetic field was applied by a flux line. A bias tee was used to divide the DC and RF currents.

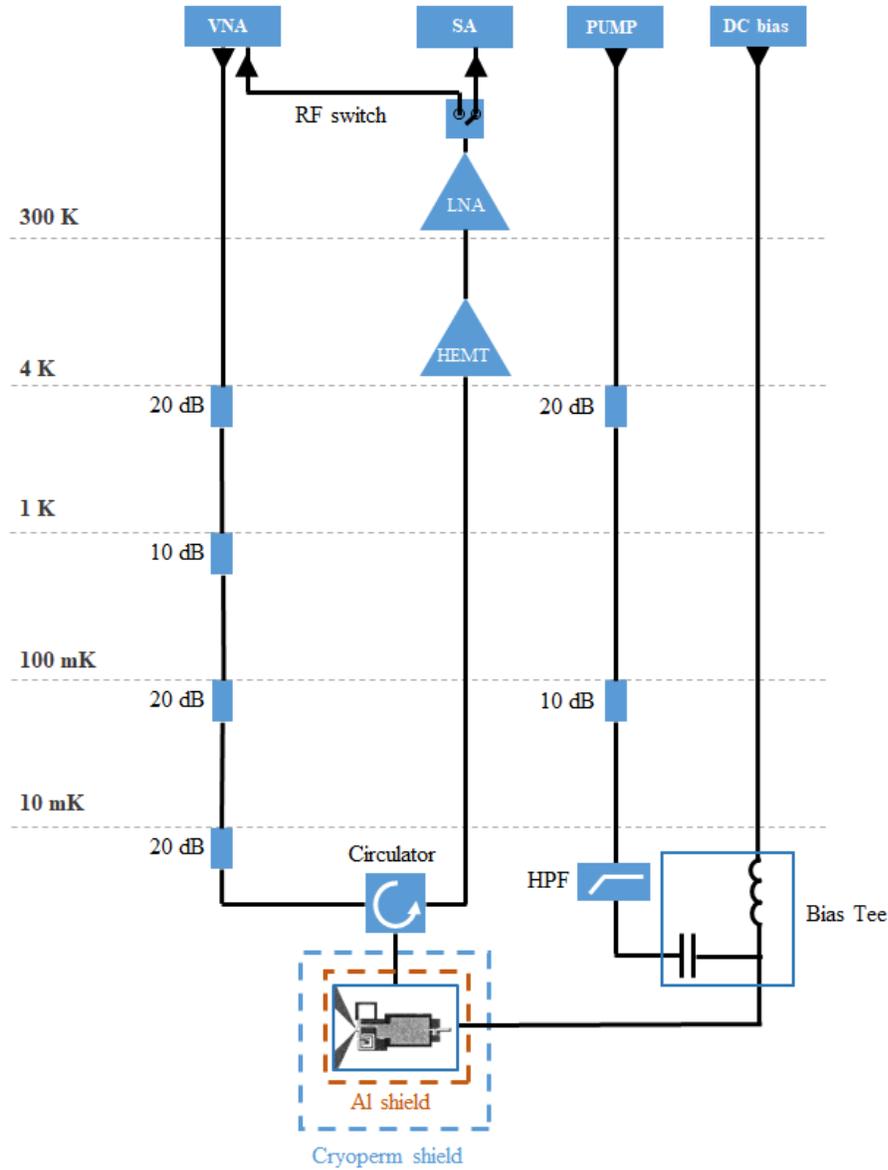

FIG. S2. Schematic of the cryogenic microwave measurement setup

**Gain and bandwidth**

The device operates in three-wave mixing parametric amplification. We designed our IMPA for the center frequency of $\omega_0/2\pi$ = 6.5 GHz, but the frequency and the gain response depend on the input line elements. Non-ideal wire-bond connections, circulators and cables lead to modifying the gain-frequency response. Therefore, first we find the optimal operating point at different pump conditions. In FIG. S3, we display the gain for various pump frequencies.

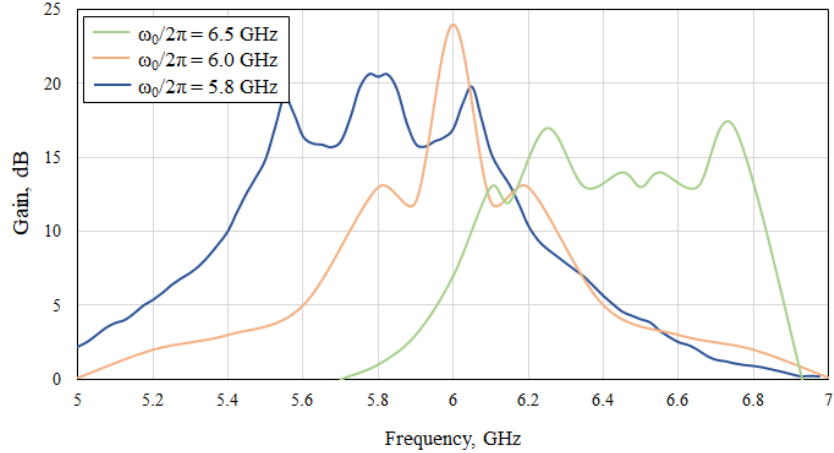

FIG. S3. Gain and bandwidth of proposed device at different pump conditions

We observed two-pole response corresponding to the two-pole Chebyshev matching network at design frequency. But the optimal gain and bandwidth response were reached at $\omega_0/2\pi = 5.8$ GHz. A three-pole response indicates of influence of input line on the IMPA characteristics. FIG. S4 show simulation results of the input line parameters and corresponding gain performance.

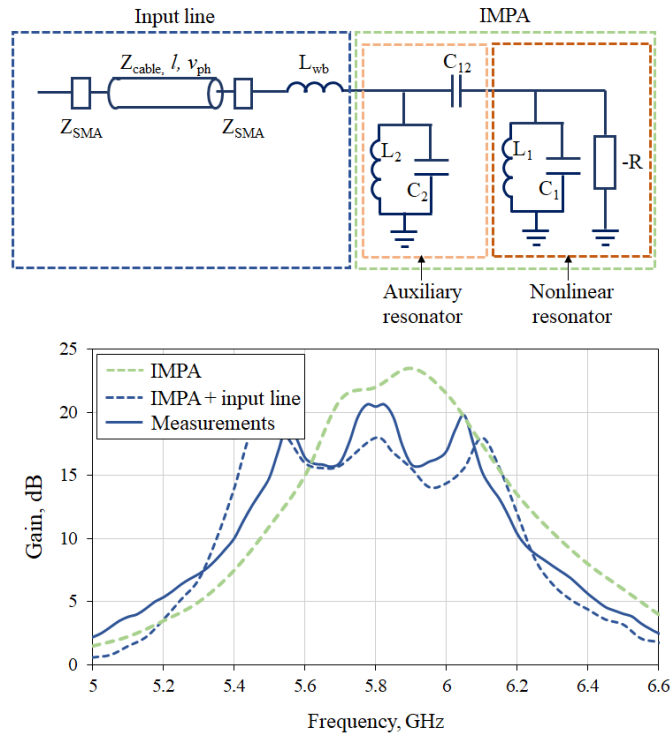

FIG. S4. Comparison between simulated and measurement of gain response

The gain response depending on input line parameters was calculated. The wire-bond connections was modeled using inductance element, and SMA connectors and distance between the IMPA and circulator were modeled using transmission line segments with characteristic impedances $Z_i$ and lengthes

$l_i$. The IMPA can be modeled as a parallel LC resonator shunted by a negative resistance with auxiliary resonator. We used the fact length of coaxial cable $l = 30$ cm, the magnitude of reflection corresponding to the SMA datasheet and wire-bond connection inductance $L_{wb} = 0.3$ nH and observed the gain response similar to measurement results.

**Engineering guide of noise temperature measurement**

The noise performance of the IMPA was characterized using the signal-to-noise ratio (SNR) improvement method, giving noise relative to the system noise temperature [5]. The amplification chain consisted of the IMPA and HEMT with a gain of 40 dB. In this method, the SNR is first measured when the amplifier is turned off. Then, the amplifier is turned on, and the SNR is measured once again. The SNR dependence on gain is defined as follows [6]:

$$\text{SNR}(G) = \frac{GP_{signal}}{G * P_{noise_{in}} + (G-1) * P_{IMPA} + P_{HEMT\_4K}} \quad (S5)$$

where G is the power gain, $P_{signal}$ and $P_{noisein}$ indicate the signal and noise power at the input of the parametric amplifier, $P_{IMPA}$ and $P_{HEMT\_4K}$ are the noise powers of the IMPA and the HEMT amplifier. SNR without amplification (when pump turned off):

$$SNR_{off} = SNR(G=1) = \frac{P_{signal}}{P_{noise_{in}} + P_{HEMT_{4K}}} \approx \frac{P_{signal}}{P_{HEMT_{4K}}}, \quad (S6)$$

where the last approximation is performed under the condition $P_{HEMT_{4K}} \gg P_{noise_{in}}$. We used this approximation since the IMPA located in the mixing plate of dilution refrigerator. Based on the equations (S5) and (S6) the system noise at the output of the IMPA is defined as follows:

$$P_{sys} = P_{noise_{in}} + \frac{(G-1)}{G} * P_{IMPA} = \frac{SNR_{off}}{SNR(G)} * P_{HEMT_{4K}} - \frac{P_{HEMT_{4K}}}{G} \quad (S7)$$

By comparing the SNR values in these two experiments and taking into account the HEMT noise temperature (which is determined from manufacturer datasheet), we evaluated the system noise temperature when IMPA was turned on (S8).

$$T_{sys} = \frac{SNR_{off}}{SNR(G)} * T_{HEMT_{4K}} - \frac{T_{HEMT_{4K}}}{G} \quad (S8)$$

Below we demonstrate an engineering guide for the noise temperature measurement:

1. Choose a set of input signal frequencies covering the bandwidth;
2. For each frequency define the gain and SNR when the pump is turned off and turned on. To collect statistics, using VNA, we performed 15000 repeated measurements of the S21 for each frequency in the set. The SNR is defined as follows:

$$\text{SNR} = \left(\frac{\langle S_{21}\rangle}{\sigma}\right)^2, \tag{S9}$$

where $\langle S_{21}\rangle$ is the average complex transmission coefficient and $\sigma$ is the standard deviation. The measurement uncertainty originates from the estimated $T_{\text{HEMT\_4K}}$ range.

3. Compute of the noise temperature for choosing frequency (S8);
4. Repeat points 2 and 3 for each frequency of the set.